\documentclass[]{aa}
\usepackage[utf8]{inputenc}
\usepackage{color}
\usepackage[usenames,dvipsnames,svgnames,table]{xcolor}
\usepackage{txfonts}
\usepackage[]{hyperref}
\hypersetup{colorlinks=true,citecolor=blue,  linkcolor=blue,urlcolor=black}

\title{A revolution is brewing: observations of TRAPPIST-1 exoplanetary system fosters a new biomarker}

 \author{M. Turbo-King \inst{1}
\and B.R. Tang \inst{1} 
\and Z. Habeertable \inst{1} 
\and M.C. Chouffe \inst{1} 
\and B. Exquisit \inst{1} 
\and L. Keg-beer \inst{1}}

 \institute{$^{1}$ La R{\'e}sonance de Laplace, Paris, France; \email{resonancedelaplace@gmail.com}}

\date{April 1st, 2017\\Accepted for publication in \emph{Journal of Astrobrewlogy}}

\abstract{The recent discovery of seven potentially habitable Earth-size planets around the ultra-cool star 
TRAPPIST-1 has further fueled the hunt for extraterrestrial life. Current methods focus on closely monitoring 
the host star to look for biomarkers in the transmission signature of exoplanet's atmosphere. However, the outcome of these methods 
remain uncertain and difficult to disentangle with abiotic alternatives.
Recent exoplanet direct imaging observations by THIRSTY, an ultra-high contrast coronagraph located in La Trappe (France), 
lead us to propose a universal and unambiguous habitability criterion which 
we directly demonstrate for the TRAPPIST-1 system. Within this new framework, 
we find that TRAPPIST-1g possesses the first unambiguously habitable environment in our galaxy, 
with a liquid water percentage that could be as large as $\sim~90~\%$. 
Our calculations hinge on a new set of biomarkers, CO$_2$ and C$_{x}$H$_{2(x+1)}$O (liquid and gaseous), that could cover up to $\sim~10~\%$ 
of the planetary surface and atmosphere.
THIRSTY and TRAPPIST recent observations accompanied by our new, unbiased habitability criterion  
may quench our thirst for the search for extraterrestrial life. However, the search for intelligence must continue within and beyond our Solar System. }

\titlerunning{A revolution is brewing}
\authorrunning{Turbo-King et al.}

\begin{document}
\maketitle

\section{Introduction}

The question of whether or not there is life in the Universe has arisen since the 
very first appearance of intelligent life on the Earth. 
It is a question that has inspired the imagination of scientists, 
artists, and philosophers, culminating in the landmark philosophical paper 
by \cite{Beck:1971}. 
Although attempts to find (and indeed, even define) life and intelligent life on Earth \citep{la1916gunshot,stout2006between} 
and elsewhere have thus far met with limited success, 
with the discovery of thousands of planets outside our Solar System, we are now poised to make great progress in this arena.


The difficult question of what is required for life to exist in the Universe can be split 
up into smaller, yet still challenging, outstanding questions as summarized in the Drake equation, 
also referred to as Drake’s Gold Alien Location Equation [Drake’s Gold ALE]. 

One of the key questions that has arisen in recent times is the boundaries of the Habitable Zone, 
classically defined as the range of orbital distances in a stellar system where surface liquid water could be stable \citep{Kasting:1993}. 
It is, however, seldom addressed how this concept might become obsolete 
in exoplanetary worlds where the priorities for living a good life are 
different than getting liquid water everywhere and at all times (Kim Kardashian, \textit{personal communication}). 
A groundbreaking paper to that respect is the study by \citet{Kane:2014} which demonstrated that the existing concepts about 
the Habitable Zone completely overlooked the risk posed by Zombie attacks.

We propose to take this line of research a step further by questioning the very concept of biomarkers. 
Most studies on biomarkers chose to focus on oxygen and methane. The former is not a good choice because abiotic oxygen might arise as result of photodissociation of water molecules \citep{Wordsworth:2014}; the latter is not a good choice because the controversial discovery of methane on Mars is still so difficult to understand that a new paper on methane is not desirable \citep{Zahnle:2011}. 

We propose instead a new biomarker completely overlooked in previous studies: beer. We argue this bio-marker is universal: everyone drinks beer and, contrary to wine and vodka, non-alcoholic beer does exist \citep{Light:1988}. Another robust piece of evidence is that beer is cheaper than water and sodas in most parts of the Earth \citep{carney:2013}.

Furthermore, perhaps the only common ground amongst nations on Earth is that all make their own beer, promoting beer as the most universal and thus reliable biomarker to trace life on Earth, and by extension elsewhere in the universe. Last but not least, beer is known for its property to generate more liquid than what was ingested \citep{Bladder:1601}, making it superior to liquid water to that respect. 

\begin{figure}
  \centering
    \includegraphics[width=0.49\textwidth]{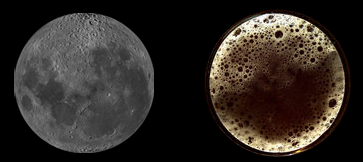}
   \caption{\emph{Direct imaging (right panel) of a beer planet using the ground-based coronagraph THIRSTY (Telescope HIgh-Resolution for Systems Transiting Yeast). This planet named THIRSTY-1664b orbits the 1664th host star detected by the THIRSTY instrument. North is up. The higher terrains seen in the northern hemisphere are informally called Sierra Nevada (note that all names are informal). The shape of craters (e.g. the giant Chimay crater) might indicate the presence of liquid beer, or at least beerthermalism. The dark plains in the southern hemisphere, named Ale Planitia, may be an evidence of ancient resurfacing by beervolcanoes (akin to the seas on the Moon). The detached layers that can be distinguished close to the northern and southern polar regions could be related to the formation of high-altitude beer clouds. By comparison, an image of the Moon (not to scale) is shown on the left pannel. \label{beerplanet}}}
\end{figure}

Like all possible alcoholic beverages, beer needs to be drunk responsibly; nevertheless, it is the human stupidity which is responsible for most beer-driven catastrophies\footnote{The best known exemple is the London Beer Flood that happened on 17 October 1814 and caused the death of 8 people.}. Unfortunately, a world devoid of beer will still be prone to the worst tendencies for racism, misogyny and homophobia. 
Sadly, stupidity might even be a better biomarker than beer; measuring this quantity remotely is however impossible, which has the fortunate consequence that ALE-iens might still want to visit us humans, and thus benefit from favorable in-situ biomarkers such as love \citep{valentine1999food}, heavy-metal music \citep{Cann:90}, and food \citep{Ramsay:1925}. Despite their desirable qualities, the latter biomarkers are not as easily detectable and abundant as beer \citep{King:01}, which remains by far the perfect biomarker in the Universe.

The technological leap forward that has been propulsed by modern astronomy now enables us to directly image planets in which the beer biomarker might not only be in gaseous phase but also in stable liquid phase at the surface. The breathtaking Figure~\ref{beerplanet}, obtained recently with the ground-based coronagraph THIRSTY (Telescope of HIgh-Resolution for Systems Transiting Yeast)\footnote{the project was initiated more than 40~years ago by \cite{Beer:71}.}, features a direct imaging of a putative beer planet in which either liquid beer \citep{Beer:16} or beervolcanism \citep{debeer:1962} could have shaped the surface of the planet. The famous THIRSTY image also features a detached layer likely to be high-altitude clouds \citep{Bohr:13} made of liquid beer droplets \citep{Beer:09}. Such breakthrough imaging opens new perspectives not only on the importance of beer as a biomarker, but also on the concept of ``Ha-beer-table Zone'' i.e. the range of orbital distances to the stellar object in which liquid beer is stable in vast areas across the surface of the planet \citep{Beer:08}, stellar variability throughout geological ages being accounted for \citep{Beer:06sun}.

\section{A quantitative assessment of the Ha-beer-table Zone}

The Ha-beer-table Zone has remained thus far a theoretical concept seldom discussed quantitatively in the literature. 
In this section, we gather all the existing criteria for the stability of liquid beer and plot, 
for the first time, the Ha-beer-table Zone diagram. It is a groundbreaking result, 
whose applications to the terrestrial environment were also validated by the authors themselves 
through \emph{in-situ} immersion in a beer jacuzzi.

\subsection{Ha-beer-table Zone criteria}

Several distinct criteria can be listed to help define the Ha-Beer-Table Zone.
\begin{description}
\item \textbf{The Homer Simpson criterion} \citep{Hint:97}: The surface temperature of the planet shall remain between 10 and 20 degrees Celsius (283 and 293 Kelvins), which is the optimal temperature to conserve beer.

\begin{table}
\centering
\caption{CO$_2$ volumes for various types of beers and associated CO$_2$ partial pressures (in bar). Adapted from the chart from Principles of Brewing Science. 1 volume of CO$_2$ corresponds to $\sim$0.045 moles (= 2g) of CO$_2$ per liter of beer.}
\resizebox{9cm}{!}{
\begin{tabular}{lcccc}
\hline
Type of beer & CO$_2$ volume & $P_{\text{CO}_2}$(10$^{\circ}$C) & $P_{\text{CO}_2}$(20$^{\circ}$C)  &  \\
\hline
British Style Ales & 1.7 & 0.75 & 1.3 \\
Porter, Stout & 2.0 & 1.0 & 1.7 \\
Belgian Ales & 2.2 & 1.2 & 1.9 \\
American Ales and Lager & 2.5 & 1.6 & 2.3 \\
Fruit Lambic & 3.7 & 2.8 & 3.8 \\
German Wheat Beer & 3.9 & 3.0 & 4.0 \\
\hline
\end{tabular} }
\label{table_beer_co2} 
\end{table}

\item \textbf{The Perrier-San Pellegrino criterion} \citep{Bado:53} To maintain one's beer carbonated, one needs to pressurize it. Therefore, any computation of the Ha-beer-table Zone shall account for the pressurization of beers of various levels of carbonation. For various kinds of beers, Table~\ref{table_beer_co2} summarizes the CO$_2$ partial pressure needed to pressurize the beer.

\item \textbf{The Clausius-Clapeyron-Henry-Budweiser criterion} \citep{Bud:71}:  For each type of beer, the partial pressure of ethanol $P_{\text{C}_2\text{H}_6 \text{O}}$ will be different in the atmosphere, and a function of both temperature and concentration of alcohol $[{\text{C}_2\text{H}_6 \text{O}}]_l$ in the beer ocean (see Table~\ref{table_beer_alcohol}):

\begin{equation}
P_{\text{C}_2\text{H}_6 \text{O}}=\frac{[{\text{C}_2\text{H}_6 \text{O}}]_l}{H_{\text{C}_2\text{H}_6 \text{O}}(T)}, \\
\label{eq_henry}
\end{equation}
with $H_{\text{C}_2\text{H}_6 \text{O}}(T)$ the Henry Weinhard's constant.

\begin{table}
\centering
\caption{Alcohol/ethanol partial pressures (in millibars) for drinks with various volumes of alcohol.}
\resizebox{9cm}{!}{
\begin{tabular}{lcccc}
\hline
Alcohol by volume & molar concentration & P$_{\text{C}_2\text{H}_6\text{O}}$(10$^{\circ}$C) &  P$_{\text{C}_2\text{H}_6\text{O}}$(20$^{\circ}$C) \\
\hline
3$^\circ$ & 0.009 & 0.36 & 0.72 \\
5$^\circ$ & 0.016 & 0.64 & 1.3 \\
7$^\circ$ & 0.023 & 0.92 & 1.8 \\
9$^\circ$ & 0.030 & 1.2 & 2.4 \\
11$^\circ$ & 0.037 & 1.5 & 3.0 \\
13$^\circ$ & 0.044 & 1.8 & 3.5 \\
\hline
\end{tabular} }
\label{table_beer_alcohol} 
\end{table}

In summary, the three previous criteria tell us that atmospheres of beer planets should be composed of 0.7-4.0~bars of CO$_2$, as well as 0.-3.5 millibars\footnote{The 0~millibar case should be disregarded here, since we definitely do not want to lose our time to look for alcohol-free beer exoplanets.} of ethanol.

\item \textbf{The Sunburn criterion} \citep{Tan:04}: UV photons are extremely harmful for the fermenting ability of yeasts \citep{Tanner:1934}, and therefore could significantly spoil the taste of the beer. Depending on the amount of CO$_2$ (e.g. the level of carbonation) in the ocean of beer, a planet could have two different ways to be self protected from UV radiation and therefore to keep a good taste:

\begin{enumerate}
\item A low carbonated beer planet must have a significant ozone layer that is very efficient to absorb UV light.
\item A highly carbonated beer planet is expected to form a complete foam cover that should naturally protect the precious beer from UV radiation.
\end{enumerate}

In any case, it has been shown, using the Beer-Lambert law \citep{Beer:1852}, that UV light should not penetrate deeper than few tens of meters in beer oceans \citep{Cockell:2000}. This also tells us that, even though it might be tempting to drink surface liquid beer when reaching (after a long journey) a distant beer planet, make no mistake, the best ale lies meters below.

\end{description}

\begin{figure*}
  \centering
    \includegraphics[width=1.0\textwidth]{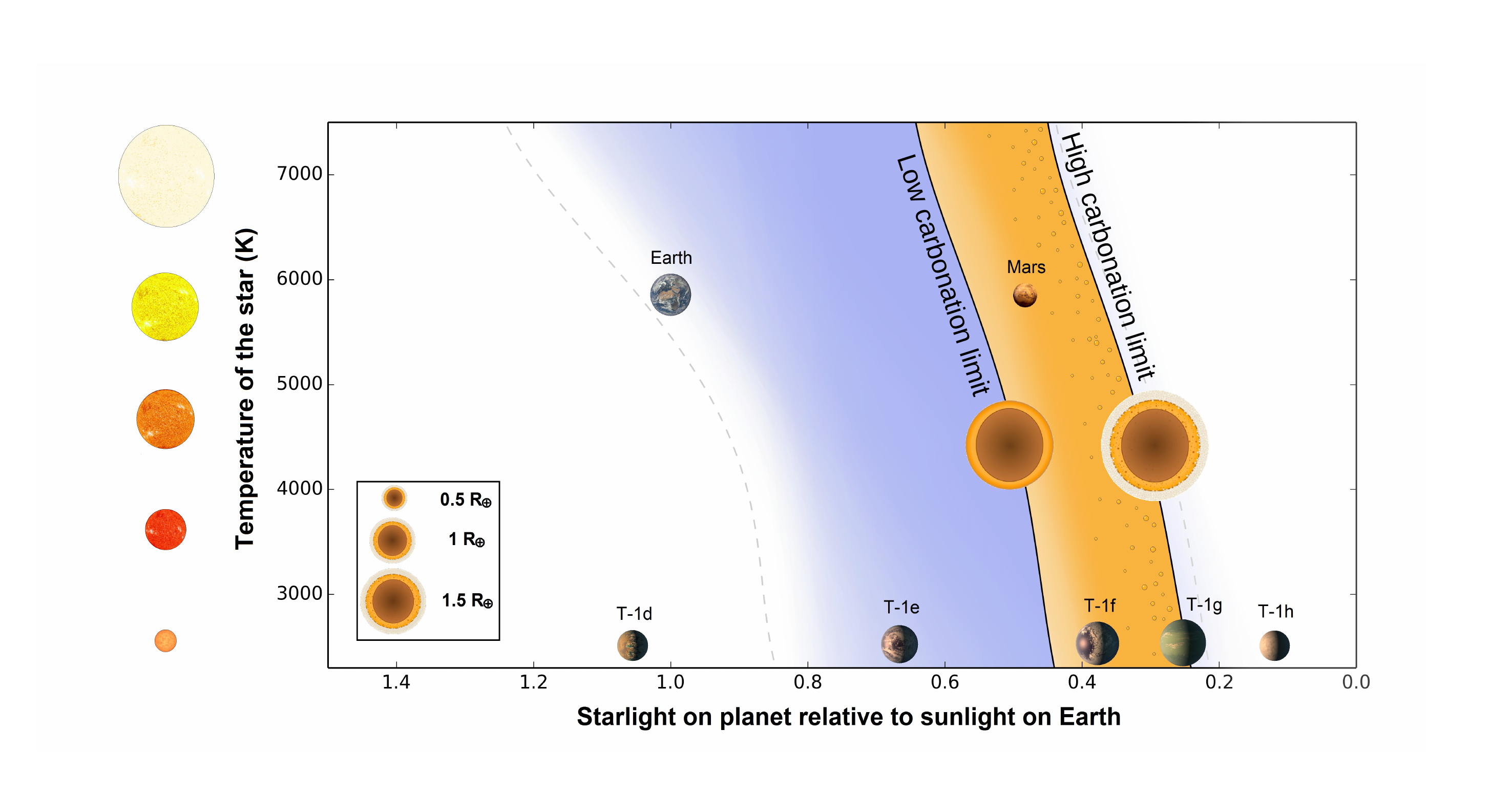}
   \caption{\emph{Diagram showing the radial extent of the Ha-beer-table Zone as a function of stellar temperature. As a comparison, dashed lines corresponds to the traditional limit of the old-fashioned Habitable Zone \citep{Kopp:2013}.}}
\label{habeertablezone_diag}
\end{figure*}

\subsection{Inner and outer edges of the Ha-beer-table Zone - Low and high carbonation limits}

Following the previous criteria, the inner edge of the Ha-beer-table Zone is set at the orbital distance for which a low-carbonated (P$_{\text{CO}_2}$~=~0.7~bar), alcohol-free beer planet has a surface temperature of 20$^{\circ}$C. 

The outer edge of the Ha-beer-table Zone is set at the orbital distance for which a highly-carbonated (P$_{\text{CO}_2}$~=~4~bars) beer planet with a high volume of alcohol (arbitrarily fixed at 11$^{\circ}$ in the present study; $P_{\text{C}_2\text{H}_6 \text{O}}$=~2~millibar) has a surface temperature of 10$^{\circ}$C. 

There are three good reasons to preferentially hunt planets near the outer edge of the Ha-beer-table Zone:
\begin{enumerate}
\item Such planets are expected to maintain high surface pressures (up to 4~bars of CO$_2$) that tend to homogenize the temperature of the ocean of beer and thus favor its conservation. This is of high importance on tidally locked beer planets.
\item Such planets would 1) receive fewer UV flux and 2) form a thick layer of foam because they are expected to be highly carbonated. These two effects combined favor the protection of the beer.
\item A trivial and highly reasonable argument is clearly to look for planets where the degree of alcohol is higher. 
\end{enumerate}

\subsection{Ha-beer-table Zone diagram}

Using the DML (Draught Micro-Labrewatory) 3-D Full Global Climate Model\footnote{The model includes a generalized radiative transfer designed for cocktails of beer atmospheres, and takes into account various processes such as the radiative effect of beer clouds and the surface albedo of foam.} designed to simulate the climate of any kind of planets around any kind of stars, we computed both the inner and outer edges of the Ha-beer-table Zone, for various types of stars and various orbital distances. Our results are summarized on Figure~\ref{habeertablezone_diag}. Implications of these results are discussed in section~\ref{habeertability}.

\subsection{Going further}

The climate of beer planets is not trivial to assess because of the many possible retroactions related to the beer surface properties.

As a first example, even weak tidal waves (induced by the star or close-in planets) could sufficiently shake the beer ocean so that a very thick layer of foam forms at its surface. This material is known to be both very porous and bright, and it is very likely that it will impact the diurnal surface temperatures and the cycles of beer condensation and sublimation. It is a challenging task to evaluate how long such a foam can subsist on the beer planet's surface with a changing thermal inertia. The stronger the waves, the thicker the foam and the longer it will remain on the surface. Further studies should also consider in detail how the presence of a thick layer of foam could affect the tidal dissipation of a beer planet and thus calculations of its orbital evolution.

In any case, significant progress could be made by laboratory experiments of:
\begin{enumerate}
\item[$\bullet$] mechanical properties of foam for various kind of beers.
\item[$\bullet$] reflexion spectrum of foam. This might be of importance around cool stars like TRAPPIST-1 as previously shown by \citet{Joshi:2012}.
\item[$\bullet$] absorption spectra of ethanol-carbon dioxide mixtures for multiple sets of temperature and pressure.
\end{enumerate}

Such measurements could be made extremely difficult because in-situ ingestion of alcohol by experimenters would significantly increase the size of the error bars.

\section{Ha-beer-tability in the Solar System and beyond}
\label{habeertability}

\subsection{Beer Planets in the Solar System}

An unexpected corollary of the groundbreaking diagram in Figure~\ref{habeertablezone_diag} is that Mars is located in the Ha-Beer-Table Zone. It might explain the recent bursts of methane detected by the Curiosity rover \citep{Webs:15}, although further work is needed to confirm this possibility. 

Fortunately, the presence of an underground yeast-rich ocean of beer on Mars might be tested by future missions of exploration on Mars like Mars 2020 or Exomars 2020.

\subsection{TRAPPIST-1 is the first system in the Ha-beer-table Zone}

After Mars, Kepler-62f, Kepler-1229b, Kepler-186f, GJ581d, TRAPPIST-1f and g are indeed the first planets in the Ha-beer-table Zone.
However, preliminary 3-D climate simulations of TRAPPIST-1f - assuming a synchronous rotation - show that even though mean surface temperatures of 10-20$^{\circ}$C could be achieved, the surface temperature is expected to reach locally temperatures as high as 40$^{\circ}$C and as low as -10$^{\circ}$C, giving the surface liquid beer an unpleasant taste.
TRAPPIST-1g could be a well carbonated, alcohol-rich planet, as supported by 1) the Time Transit Variation (TTV\footnote{not to be confused with Time Turbidity Variation.}) analysis of the planet \citep{Gillon:2017} and 2) the fact that TRAPPIST-1 planets are today in a near-resonant chain, suggesting that the planets could have formed far from their star (beyond the beer-line) and migrated afterward to their current position. TRAPPIST-1g is therefore the best known candidate for ha-beer-tability outside our Solar System and 100$\%$ of James Webb Space Telescope (JWST)\footnote{even between two transits.} observing time should be dedicated to characterizing the atmosphere of this potentially ha-beer-table planet. 

Although we know from history that giving a name to a planet increases the probability that it does not exist (e.g. \citet{Vogt:2010}), we cannot help but propose 7 names that should be adopted in the future for the 7 wonders of the TRAPPIST-1 system:

\begin{itemize}

\item[$\bullet$] \textbf{TRAPPIST-1b as \textit{Achel the brown}.}
\item[$\bullet$] \textbf{TRAPPIST-1c as \textit{Isid'or the amber}.}
\item[$\bullet$] \textbf{TRAPPIST-1d as \textit{Trappe the blond}.}
\item[$\bullet$] \textbf{TRAPPIST-1e as \textit{Chimay la bleue}.}
\item[$\bullet$] \textbf{TRAPPIST-1f as \textit{Chimay la rouge}.}
\item[$\bullet$] \textbf{TRAPPIST-1g as \textit{Chimay the gold-i-locked}.}
\item[$\bullet$] \textbf{TRAPPIST-1h as \textit{Chimay la blanche}.}
\end{itemize}

\subsection{Beeruptions and implications}

We examined the consequences of tidal heating for TRAPPIST-1 planets.
Following Peale et al. 1979 one can predict the heating, and eventual extreme beervolcanism due to forced eccentricities. 
Assuming an Io-like rigidity $\mu = 6.5 \cdot 10^{11} $ dynes cm$^{-2}$, and a tidal heating factor $Q \sim 100 $, 
we calculate that \textit{Chimay la rouge} and \textit{the gold-i-locked} - the two potentially habeertable planets of TRAPPIST-1 system - 
could experience reccurent beeruptions. If these events occur during transits, they could transiently and significantly 
increase the transit depth of the two planets. 
Even though preliminar analysis using Kepler/K2 data did not show any sign of beervolcanism activity \citep{Luger:2017}, such detection could be 
attempted by rigorous follow-up observations of TRAPPIST-1 light curve.

\section{Beer: one biomarker to rule them all}

\begin{figure}
  \centering
    \includegraphics[width=0.5\textwidth]{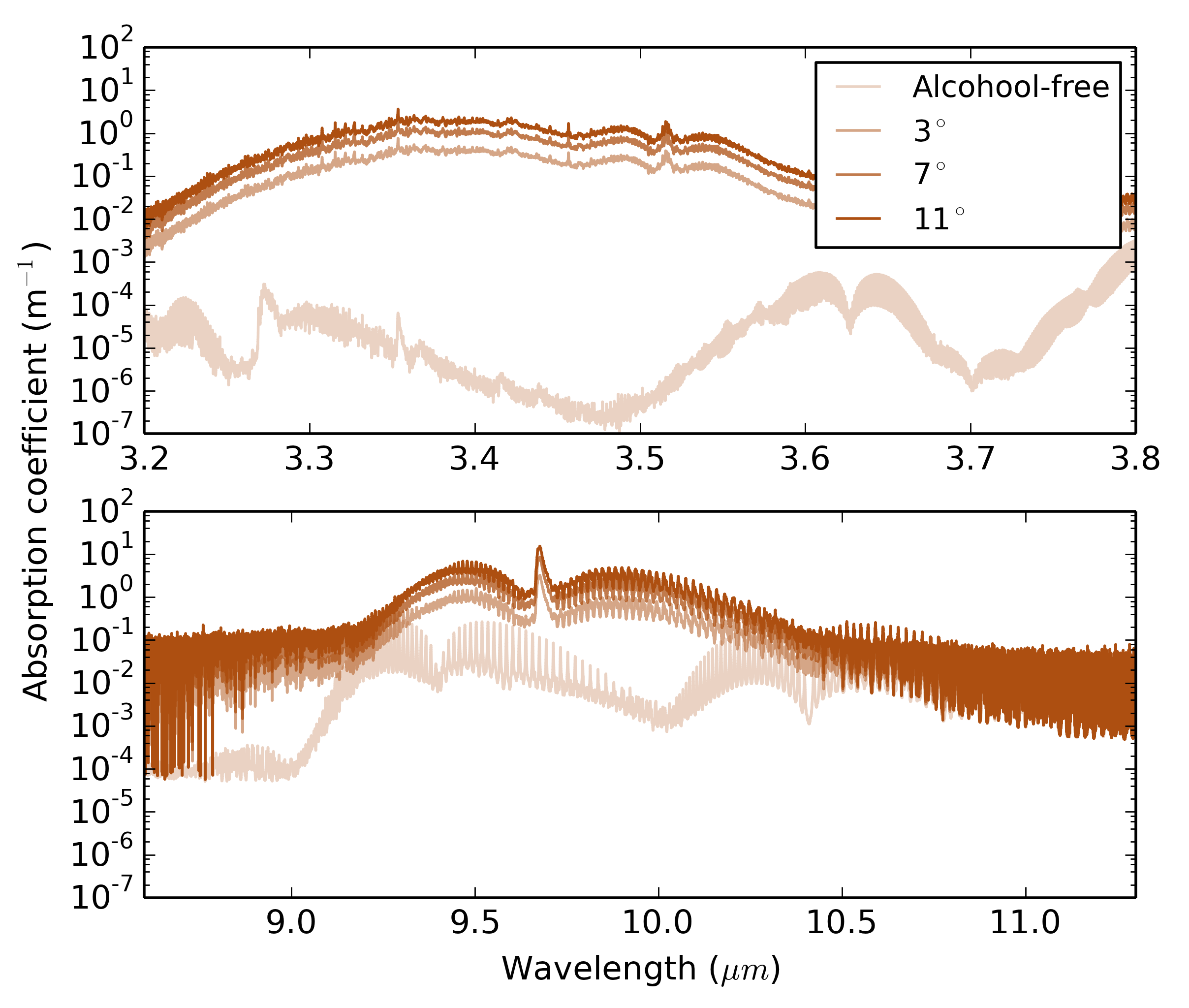}
   \caption{\emph{Synthetic infrared absorption spectrum of a beer planet, at P~=~1~bar and T~=~293K, for beer planets at several degrees of alcohol, and compared with a pure CO$_2$ (alcohol-free) atmosphere. Alcohol vapor and byproducts cause a significant rise of opacity from 3 to 4 microns, and from 8.5 to 10.5 microns. Our calculations use the HITRAN database \citep{hitran2012}. Lack of data for ethanol forced us to use absorption lines of an excellent proxi: methanol \citep{HITRAN-880}. We take therefore this opportunity to encourage the HITRAN consortium to set up absorption lines of ethanol. In any case, we looked for interesting spectral features at 1664~cm$^{-1}$ but did not find any.}}
\label{synth_spectrum}
\end{figure}

A few bars of CO$_2$, a few millibars of alcohol, a cover of foam or an ozone layer; that's all you need to build the atmosphere of a beer planet. We computed the synthetic\footnote{Note that, although the spectrum is called "synthetic", the beer should be preferentially organic.} spectrum of such atmosphere (see Figure~\ref{synth_spectrum}) and found many interesting features that could be used to identify remotely beer exoplanets:

\begin{enumerate}
\item[$\bullet$] Ethanol/methanol are very strong absorbents around 3-to-4 and 8.5-to-10.5 microns. Even light beer oceans are expected to generate atmosphere extremely opaque in the regions where ethanol and methanol absorb. Ethanol (in particular when associated to a gas like CO$_2$) is in fact a powerful greenhouse. This plays an important role in the calculation of the edges of the Ha-beer-table Zone (see Figure~\ref{habeertablezone_diag}).
\item[$\bullet$] Simultaneous measurements of 1) a strong CO$_2$ absorption (in particular around 2.8, 4.3 and 15~microns) and 2) a strong ethanol absorption (around 3.5 and 9.5~microns) would be an excellent biomarker to detect beer planets. In fact, given the high opacities of ethanol and carbon dioxide in their respective domains of absorption, it would be easy (assuming low flattening by beer clouds in the spectra) to characterize the presence and volume of alcohol of an ocean of beer on TRAPPIST-1 planets using JWST. 
\item[$\bullet$] Quantifying precisely these absorptions could not only reveal the existence of a beer planet, but also tell us the degree of alcohol in it. The link might not be straightforward though, because the amount of ethanol in the upper atmosphere might be limited by the saturation. This could bias the measurement of the degree of alcohol in the beer planet. Such processes could be addressed in future studies.
\item[$\bullet$] At any rate, an ethanol mixing ratio higher than few percents would indicate the presence of rhum/vodka/... at the surface of the planet. These scenarios are out of the scope of this paper but should be addressed as a high-priority science goal.
\end{enumerate}

\section{Concluding remarks}

In this study, we calculated the boundaries of the so called Ha-beer-table Zone. We showed that planetary ha-beer-tability is critically dependent on the carbonation level of a potential beer ocean, as well as its degree of alcohol. From this definition, we derived a new combination of biomarkers that should be used in the future - in particular with JWST through transit spectroscopy - to detect and characterize beer planets.

We know that Earth-like planets, Jupiler-like planets... are common in our galaxy and by extension in the Universe. Following the discoveries of the TRAPPIST-1 planets, and given the large extent of the Ha-beer-table zone as calculated in our study, beer planets should also be extremely common in our galactic neighborhood. 

More generally, our work raises the question of how extraterrestrial life would evolve on beer planets. At first sight, beer should be a solvent at least as good as water for life to emerge \citep{Miller:1953}. However, we are worried about what form would take natural selection on such planets and if drunk living organisms would really be able to evolve in what we - humans - call the intelligent life paradigm. If beer planets are really common, this is an alternative scenario that should be assessed to solve the Fermi paradox.




\section{Acknowledgements}

We thank Belgian monks and astronomers for inspiring this work. This project has received funding from the Agence Nationale de l’Excellence Scientifique (ANES) under 
the Emerge $\&$ See research and innovation programme (grant agreement No. 016640).

\bibliographystyle{apalike} 
\bibliography{biblio} 

\begin{thebibliography}{}

\bibitem[Badoit et~al., 1953]{Bado:53}
Badoit, J., Perrier, J., and San~Pellegrino, M. (1953).
\newblock The co$_2$ content of drinks, except champagne.
\newblock {\em Journal of Fluid Mechanics}.

\bibitem[Beck, 1971]{Beck:1971}
Beck, L.~W. (1971).
\newblock Extraterrestrial intelligent life.
\newblock {\em Proceedings and Addresses of the American Philosophical
  Association}, 45:5--21.

\bibitem[Beer, 1852]{Beer:1852}
Beer (1852).
\newblock Bestimmung der absorption des rothen lichts in farbigen
  flüssigkeiten.
\newblock {\em Annalen der Physik}, 162(5):78--88.

\bibitem[Beer, 2016]{Beer:16}
Beer, A. (2016).
\newblock {\em Vistas in Astronomy: Solar-Terrestrial Relations, Geophysics,
  Planetary System, Stellar Astronomy, Photometry, Spectroscopy, Spectral
  Peculiarities and Novae, Galaxies, Cosmogony and Cosmology}.
\newblock Elsevier.

\bibitem[Beer et~al., 2008]{Beer:08}
Beer, J., Abreu, J., and Steinhilber, F. (2008).
\newblock Sun and planets from a climate point of view.
\newblock {\em Proceedings of the International Astronomical Union},
  4(S257):29--43.

\bibitem[Beer et~al., 2006]{Beer:06sun}
Beer, J., Vonmoos, M., and Muscheler, R. (2006).
\newblock Solar variability over the past several millennia.
\newblock {\em Space science reviews}, 125(1-4):67--79.

\bibitem[Beer et~al., 2009]{Beer:09}
Beer, N.~R., Rose, K.~A., and Kennedy, I.~M. (2009).
\newblock Monodisperse droplet generation and rapid trapping for single
  molecule detection and reaction kinetics measurement.
\newblock {\em Lab on a Chip}, 9(6):841--844.

\bibitem[Beer et~al., 1971]{Beer:71}
Beer, R., Norton, R.~H., and Seaman, C.~H. (1971).
\newblock Astronomical infrared spectroscopy with a connes-type interferometer.
  i. instrumental.
\newblock {\em Review of Scientific Instruments}, 42(10):1393--1403.

\bibitem[Bladder, 1601]{Bladder:1601}
Bladder, P. (1601).
\newblock When the bladder is full of beer, the end is near.
\newblock {\em The legend of Tycho Brahe}.

\bibitem[Bohren, 1987]{Bohr:13}
Bohren, C.~F. (1987).
\newblock {\em Clouds in a glass of beer: simple experiments in atmospheric
  physics}.
\newblock Wiley.

\bibitem[Bud et~al., 1971]{Bud:71}
Bud, C.~G., Advisor, A., and Advisor, B. (1971).
\newblock Process for producing beer.
\newblock US Patent 3,573,928.

\bibitem[Cannibal and Corpse, 1990]{Cann:90}
Cannibal, T. and Corpse, J. (1990).
\newblock Eaten back to life.
\newblock {\em Metal Blade Records}.

\bibitem[{Carney}, 2013]{carney:2013}
{Carney}, S. (2013).
\newblock {Brewing Controversy Over Proposal to Make Water Cheaper Than Beer}.
\newblock {\em The Wall Street Journal}, 24:183--183.

\bibitem[Cockell, 2000]{Cockell:2000}
Cockell, C.~S. (2000).
\newblock Ultraviolet radiation and the photobiology of earth's early oceans.
\newblock {\em Origins of life and evolution of the biosphere}, 30(5):467--500.

\bibitem[{De Beer}, 1962]{debeer:1962}
{De Beer}, G. (1962).
\newblock The volcanoes of auvergne.
\newblock {\em Annals of Science}, 18(1):49--61.

\bibitem[De~Garde, 1916]{la1916gunshot}
De~Garde, L.~A. (1916).
\newblock {\em Gunshot injuries: how they are inflicted, their complications
  and treatment}.
\newblock W. Wood.

\bibitem[{Gillon} et~al., 2017]{Gillon:2017}
{Gillon}, M., {Triaud}, A.~H.~M.~J., {Demory}, B.-O., {Jehin}, E., {Agol}, E.,
  {Deck}, K.~M., {Lederer}, S.~M., {de Wit}, J., {Burdanov}, A., {Ingalls},
  J.~G., {Bolmont}, E., {Leconte}, J., {Raymond}, S.~N., {Selsis}, F.,
  {Turbet}, M., {Barkaoui}, K., {Burgasser}, A., {Burleigh}, M.~R., {Carey},
  S.~J., {Chaushev}, A., {Copperwheat}, C.~M., {Delrez}, L., {Fernandes},
  C.~S., {Holdsworth}, D.~L., {Kotze}, E.~J., {Van Grootel}, V., {Almleaky},
  Y., {Benkhaldoun}, Z., {Magain}, P., and {Queloz}, D. (2017).
\newblock {Seven temperate terrestrial planets around the nearby ultracool
  dwarf star TRAPPIST-1}.
\newblock {\em Nature}, 542:456--460.

\bibitem[Harrison et~al., 2012]{HITRAN-880}
Harrison, J.~J., Allen, N. D.~C., and Bernath, P.~F. (2012).
\newblock Infrared absorption cross sections for methanol.
\newblock {\em Journal Of Quantitative Spectroscopy and Radiative Transfer},
  113:2189--2196.

\bibitem[Hintz, 1997]{Hint:97}
Hintz, L. (1997).
\newblock Is the beer really better, drunk by your idol-the duff beer case.
\newblock {\em Sydney L. Rev.}, 19:114.

\bibitem[{Joshi} and {Haberle}, 2012]{Joshi:2012}
{Joshi}, M.~M. and {Haberle}, R.~M. (2012).
\newblock {Suppression of the Water Ice and Snow Albedo Feedback on Planets
  Orbiting Red Dwarf Stars and the Subsequent Widening of the Habitable Zone}.
\newblock {\em Astrobiology}, 12:3--8.

\bibitem[{Kane} and {Zelsiz}, 2014]{Kane:2014}
{Kane}, S.~R. and {Zelsiz}, F. (2014).
\newblock {A Necro-Biological Explanation for the Fermi Paradox}.
\newblock {\em ArXiv e-prints}.

\bibitem[{Kasting} et~al., 1993]{Kasting:1993}
{Kasting}, J.~F., {Whitmire}, D.~P., and {Reynolds}, R.~T. (1993).
\newblock {Habitable Zones around Main Sequence Stars}.
\newblock {\em Icarus}, 101:108--128.

\bibitem[Kingfisher and Cobra, 2001]{King:01}
Kingfisher, A. and Cobra, A. (2001).
\newblock The abundance of beer in the far-reaching universe.
\newblock {\em Astronomy and Astrophysics}.

\bibitem[{Kopparapu} et~al., 2013]{Kopp:2013}
{Kopparapu}, R.~K., {Ramirez}, R., {Kasting}, J.~F., {Eymet}, V., {Robinson},
  T.~D., {Mahadevan}, S., {Terrien}, R.~C., {Domagal-Goldman}, S., {Meadows},
  V., and {Deshpande}, R. (2013).
\newblock {Habitable Zones around Main-sequence Stars: New Estimates}.
\newblock {\em \apj}, 765:131.

\bibitem[Light, 1988]{Light:1988}
Light, W. (1988).
\newblock Production of low alcoholic content beverages.
\newblock US Patent 4,717,482.

\bibitem[{Luger} et~al., 2017]{Luger:2017}
{Luger}, R., {Sestovic}, M., {Kruse}, E., {Grimm}, S.~L., {Demory}, B.-O.,
  {Agol}, E., {Bolmont}, E., {Fabrycky}, D., {Fernandes}, C.~S., {Van Grootel},
  V., {Burgasser}, A., {Gillon}, M., {Ingalls}, J.~G., {Jehin}, E., {Raymond},
  S.~N., {Selsis}, F., {Triaud}, A.~H.~M.~J., {Barclay}, T., {Barentsen}, G.,
  {Delrez}, L., {de Wit}, J., {Foreman-Mackey}, D., {Holdsworth}, D.~L.,
  {Leconte}, J., {Lederer}, S., {Turbet}, M., {Almleaky}, Y., {Benkhaldoun},
  Z., {Magain}, P., {Morris}, B., {Heng}, K., and {Queloz}, D. (2017).
\newblock {A terrestrial-sized exoplanet at the snow line of TRAPPIST-1}.
\newblock {\em ArXiv e-prints}.

\bibitem[{Miller}, 1953]{Miller:1953}
{Miller}, S.~L. (1953).
\newblock {A Production of Amino Acids under Possible Primitive Earth
  Conditions}.
\newblock {\em Science}, 117:528--529.

\bibitem[{Ramsay} et~al., 1925]{Ramsay:1925}
{Ramsay}, G. et~al. (1925).
\newblock An epidemic of scarlet fever spread by ice cream.
\newblock {\em American Journal of Hygiene}, 5:669--81.

\bibitem[Rothman et~al., 2013]{hitran2012}
Rothman, L.~S., Gordon, I.~E., Babikov, Y., Barbe, A., Benner, D.~C., Bernath,
  P.~F., Birk, M., Bizzocchi, L., Boudon, V., Brown, L.~R., Campargue, A.,
  Chance, K., Cohen, E.~A., Coudert, L.~H., Devi, V.~M., Drouin, B.~J., Fayt,
  A., Flaud, J.-M., Gamache, R.~R., Harrison, J.~J., Hartmann, J.-M., Hill, C.,
  Hodges, J.~T., Jacquemart, D., Jolly, A., Lamouroux, J., LeRoy, R.~J., Li,
  G., Long, D.~A., Lyulin, O., Mackie, C., Massie, S.~T., Mikhailenko, S.,
  Muller, H.~S., Naumenko, O., Nikitin, A., Orphal, J., Perevalov, V.~I.,
  Perrin, A., Polovtseva, E.~R., Richard, C., Smith, M. A.~H., Starikova, E.,
  Sung, K., Tashkun, S., Tennyson, J., Toon, G.~C., Tyuterev, V.~G., and
  Wagner, G. (2013).
\newblock The hitran 2012 molecular spectroscopic database.
\newblock {\em Journal of Quantitative Spectroscopy and Radiative Transfer},
  130:4--50.

\bibitem[Stout, 2006]{stout2006between}
Stout, J.~P. (2006).
\newblock Between two wars in a breaking world.
\newblock {\em Cather Studies, Volume 6: History, Memory, and War}, page~70.

\bibitem[Tan and Siebert, 2004]{Tan:04}
Tan, Y. and Siebert, K.~J. (2004).
\newblock Quantitative structure- activity relationship modeling of alcohol,
  ester, aldehyde, and ketone flavor thresholds in beer from molecular
  features.
\newblock {\em Journal of agricultural and food chemistry}, 52(10):3057--3064.

\bibitem[Tanner and Byerley, 1934]{Tanner:1934}
Tanner, F.~W. and Byerley, J.~R. (1934).
\newblock The effect of ultraviolet light on the fermenting ability of yeasts.
\newblock {\em Archiv f{\"u}r Mikrobiologie}, 5(1):349--357.

\bibitem[Valentine, 1999]{valentine1999food}
Valentine, G. (1999).
\newblock Food, leisure and the negotiation of sexual relations.
\newblock {\em Leisure/tourism geographies: Practices and geographical
  knowledge}, 3:164.

\bibitem[{Vogt} et~al., 2010]{Vogt:2010}
{Vogt}, S.~S., {Butler}, R.~P., {Rivera}, E.~J., {Haghighipour}, N., {Henry},
  G.~W., and {Williamson}, M.~H. (2010).
\newblock The lick-carnegie exoplanet survey: A 3.1 m$_{\text{earth}}$ planet
  in the habitable zone of the nearby m3v star gliese 581.
\newblock {\em The Astrophysical Journal Letters}, 723:954--965.

\bibitem[Webster et~al., 2015]{Webs:15}
Webster, C.~R., Mahaffy, P.~R., Atreya, S.~K., Flesch, G.~J., Mischna, M.~A.,
  Meslin, P.-Y., Farley, K.~A., Conrad, P.~G., Christensen, L.~E., Pavlov,
  A.~A., et~al. (2015).
\newblock Mars methane detection and variability at gale crater.
\newblock {\em Science}, 347(6220):415--417.

\bibitem[{Wordsworth} and {Pierrehumbert}, 2014]{Wordsworth:2014}
{Wordsworth}, R. and {Pierrehumbert}, R. (2014).
\newblock {Abiotic Oxygen-dominated Atmospheres on Terrestrial Habitable Zone
  Planets}.
\newblock {\em The Astrophysical Journal Letters}, 785:L20.

\bibitem[{Zahnle} et~al., 2011]{Zahnle:2011}
{Zahnle}, K., {Freedman}, R.~S., and {Catling}, D.~C. (2011).
\newblock {Is there methane on Mars?}
\newblock {\em Icarus}, 212:493--503.

\end{thebibliography}

\end{document}